\documentclass[conference]{IEEEtran}
\usepackage[latin9]{inputenc}
\usepackage{amsmath}
\usepackage{amsthm}
\usepackage{stackrel}
\usepackage{graphicx}
\usepackage{esint}

\makeatletter
  \theoremstyle{plain}
  \newtheorem{prop}{\protect\propositionname}
  \theoremstyle{plain}
  \newtheorem{lem}{\protect\lemmaname}
  \theoremstyle{remark}
  \newtheorem{rem}{\protect\remarkname}
  \theoremstyle{plain}
  \newtheorem{thm}{\protect\theoremname}
  \theoremstyle{definition}
  \newtheorem{defn}{\protect\definitionname}
  \theoremstyle{plain}
  \newtheorem{cor}{\protect\corollaryname}

\usepackage{amsmath}
\usepackage{amsthm}
\usepackage{amssymb}
\usepackage{geometry}
\IEEEoverridecommandlockouts

\@ifundefined{showcaptionsetup}{}{%
 \PassOptionsToPackage{caption=false}{subfig}}
\usepackage{subfig}
\makeatother

\providecommand{\corollaryname}{Corollary}
\providecommand{\definitionname}{Definition}
\providecommand{\lemmaname}{Lemma}
\providecommand{\propositionname}{Proposition}
\providecommand{\remarkname}{Remark}
\providecommand{\theoremname}{Theorem}

\begin{document}
\title{
Optimal Timing in Dynamic and Robust Attacker Engagement During Advanced Persistent Threats 
}


\author{
\IEEEauthorblockN{Jeffrey Pawlick\\} 
  \IEEEauthorblockA{NYU Tandon School of Eng. and\\US Army Research Lab\\Adelphi, MD, USA\\jpawlick@nyu.edu} 
\and 
\IEEEauthorblockN{Thi Thu Hang Nguyen\\} 
  \IEEEauthorblockA{LAAS-CNRS\\7 Avenue du Colonel Roche\\31077 Toulouse, France\\tthnguye@laas.fr} 
\and
\IEEEauthorblockN{Edward Colbert\\} 
  \IEEEauthorblockA{US Army Research Lab\\and Virginia Tech\\Arlington, VA, USA\\ecolbert@vt.edu} 
\and
\IEEEauthorblockN{Quanyan Zhu\\} 
  \IEEEauthorblockA{NYU Tandon School of Eng.\\5 MetroTech Center\\Brooklyn, NY, USA\\quanyan.zhu@nyu.edu
\thanks{This work is partially supported by an NSF IGERT grant through the Center for Interdisciplinary Studies in Security and Privacy (CRISSP) at New York University, by the grant CNS-1544782, EFRI-1441140, and SES-1541164 from National Science Foundation (NSF) and DE-NE0008571 from the Department of Energy.}
\thanks{Research was sponsored by the Army Research Laboratory and was accomplished under Cooperative Agreement Number W911NF-17-2-0104. The views and conclusions contained in this document are those of the authors and should not be interpreted as representing the official policies, either expressed or implied, of the Army Research Laboratory or the U.S. Government. The U.S. Government is authorized to reproduce and distribute reprints for Government purposes notwithstanding any copyright notation herein.}}
}
\maketitle
\begin{abstract}
Advanced persistent threats (APTs) are stealthy attacks which make
use of social engineering and deception to give adversaries insider
access to networked systems. Against APTs, active defense technologies
aim to create and exploit information asymmetry for defenders. In
this paper, we study a scenario in which a powerful defender uses
honeynets for active defense in order to observe an attacker who has
penetrated the network. Rather than immediately eject the attacker,
the defender may elect to gather information. We introduce an undiscounted,
infinite-horizon Markov decision process on a continuous state space
in order to model the defender's problem. We find a threshold of information
that the defender should gather about the attacker before ejecting
him. Then we study the robustness of this policy using a Stackelberg
game. Finally, we simulate the policy for a conceptual network. Our
results provide a quantitative foundation for studying optimal timing
for attacker engagement in network defense. \end{abstract}


\section{Introduction}

\label{sec:intro}

Traditional cybersecurity techniques such as firewall defense and
role-based access control have been shown to be insufficient against
advanced and persistent threats (APTs). Recent breaches of the Democratic
National Committee \cite{stokelwalker2017hunting} and the U.S. Office
of Personal Management \cite{barrett2015opm} have highlighted that
advanced actors are capable of undermining these defenses through
social engineering, zero-day exploits, and deceptively mimicking benign
code. Intruders establish themselves with a network using techniques
such as spear-phishing or direct physical access. Bring your own device
(BYOD) aspects of wireless networks expose additional routes for malware
entry \cite{miller2012byod}. After entry, attackers move laterally
within the network to escalate privileges and advance towards a target
asset.

\subsection{Active Cyber Defense and Honeynets}

\begin{figure}
\begin{centering}
\includegraphics[width=0.8\columnwidth]{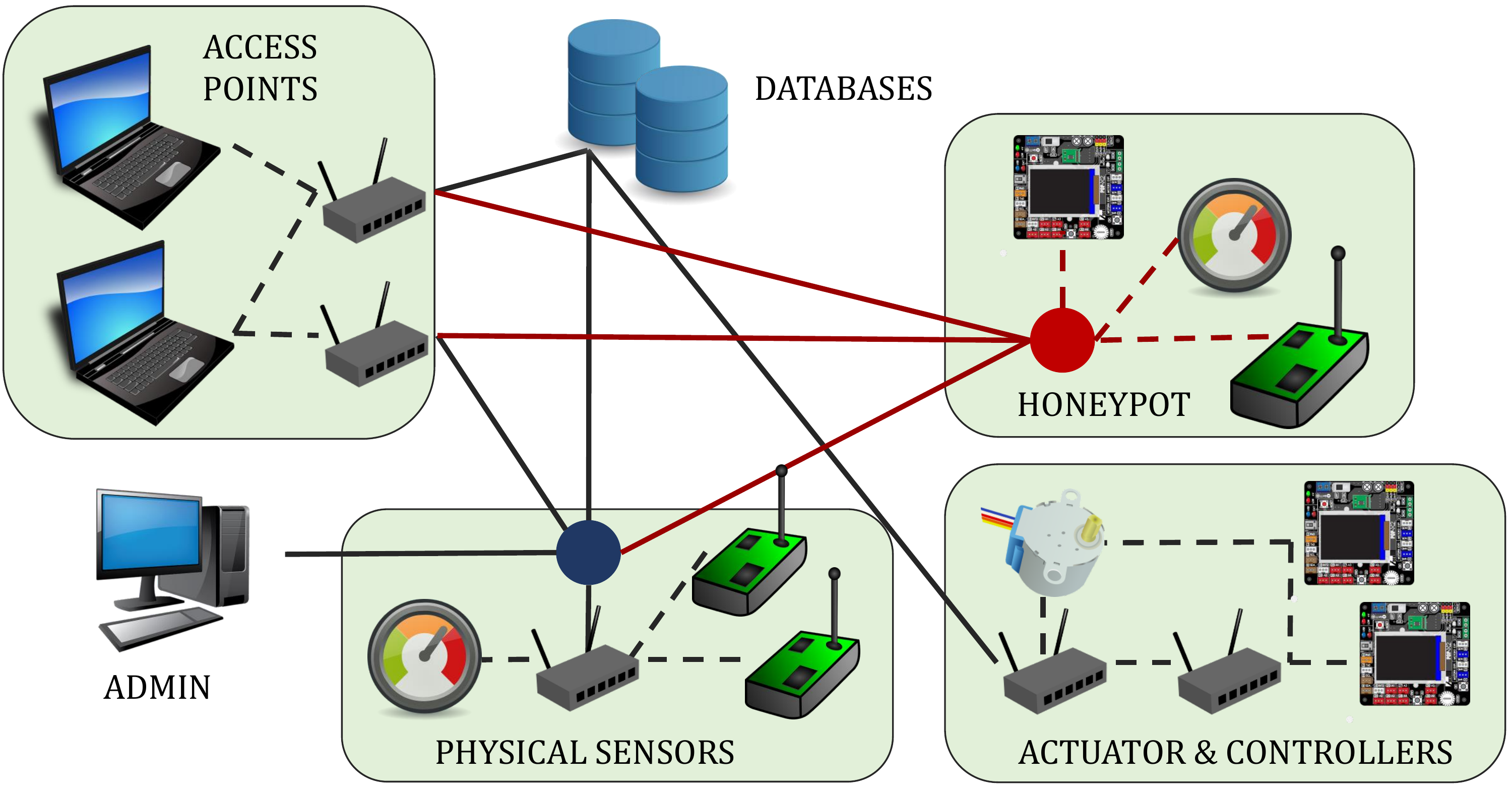} 
\par\end{centering}

\caption{\label{fig:honeyNetwork}A honeynet employed in a process control
network. Dashed (solid) lines represent wireless (wired) connections.
At the top right, a honeynet disguised as a set of sensors and controllers
records activity in order to learn about attackers.}
\end{figure}
Often security research studies deceptive attackers and purely reactive
defenders. But new techniques aim to allow defenders to gain the upper
hand in information asymmetry. The U.S. Department of Defense has
defined \emph{active cyber defense} as ``synchronized, real-time
capability to discover, detect, analyze, and mitigate threats and
vulnerabilities... using sensors, software, and intelligence...''
\cite{dod2011strategy}. These techniques both investigate attackers
and manipulate their beliefs \cite{stech2016integrating}.\emph{ Honeynets}
and \emph{virtual attack surfaces }are emerging techniques which accomplish
both purposes. They create false network views in order to lure the
attacker into a designated part of a network where he can be contained
and observed within a controlled environment \cite{albanese2016deceiving}.
Figure \ref{fig:honeyNetwork} gives a conceptual example of a honeynet
placed within a process control network in critical infrastructure
or a SCADA\footnote{Supervisory Control and Data Acquisition} system.
A wired backbone connects wireless routers that serve sensors, actuators,
controllers, and access points. A honeynet emulates a set of sensors
and controllers and records attacker activities. Engaging with an
attacker in order to gather information allows defenders to update
their threat models and develop more effective defenses.

\subsection{Timing in Attacker Engagement}

Our work considers this seldom studied case of a powerful defender
who observes multiple attacker movements within a network. This sustained
engagement with an attacker comes at the risk of added exposure. The
situation gives rise to an interesting trade-off between information
gathering and short-term security. How long should administrators
allow an attacker to remain in a honeypot before ejecting the attacker?
How long should they attempt to lure an attacker from an operational
system to a honeypot? Our abstracts away from network topology or
protocol in order to focus exclusively on these questions of timing
in attacker engagement.

\subsection{Contributions}

We make the following principle contributions: 
\begin{enumerate}
\item We introduce an undiscounted, infinite-horizon Markov decision process
(MDP) on a continuous state space to model attacker  movement constrained
by a defender who can eject the attacker from the network at any time,
or allow him to remain in the network in order to gather information. 
\item We analytically obtain the value function and optimal policy for the
defender, and verify these numerically. 
\item To test the robustness of the optimal policy, we develop a zero-sum,
Stackelberg game model in which the attacker leads by choosing a parameter
of the game. We obtain a worst-case bound on the defender's utility. 
\item We use simulations to illustrate the optimal policy for a conceptual
network.
\end{enumerate}

\subsection{Related Work}

Game-theoretic design of honeypot deployment has been an active research
area. Signaling games are used to model attacker beliefs about honeypots
in \cite{Carroll2011,pawlick2015deception}. Honeynet deployment from
a network point of view is systematized in \cite{albanese2016deceiving}.
Ref. \cite{noureddine2016game} develops a model for lateral movements
and formulates a game by which an automated defense agent protects
a network asset. Durkota et al. model dynamic attacker engagement
using attack graphs and a MDP \cite{durkota2015optimal}. Zhuang et
al. study security investment and deception using a multiple round
signaling game \cite{zhuang2010modeling}. Our work fits within the
context of these papers, but we focus on questions of timing. Other
recent work has studied timing for more general interactions in cyber-physical
systems \cite{pawlick2015flip,pawlick2017Trust} and network security
in general \cite{vanDijk2013Flip}. On the contrary, we focus on timing
in attacker engagement. Finally, this paper fits within the general
category of optimal stopping problems. Optimal stopping problems with
a finite horizon can be solved directly by dynamic programming, but
our problem has an infinite horizon (and is undiscounted).

\section{Problem Formulation}

\label{sec:modelDef}
\begin{figure}
\begin{centering}
\includegraphics[width=0.75\columnwidth]{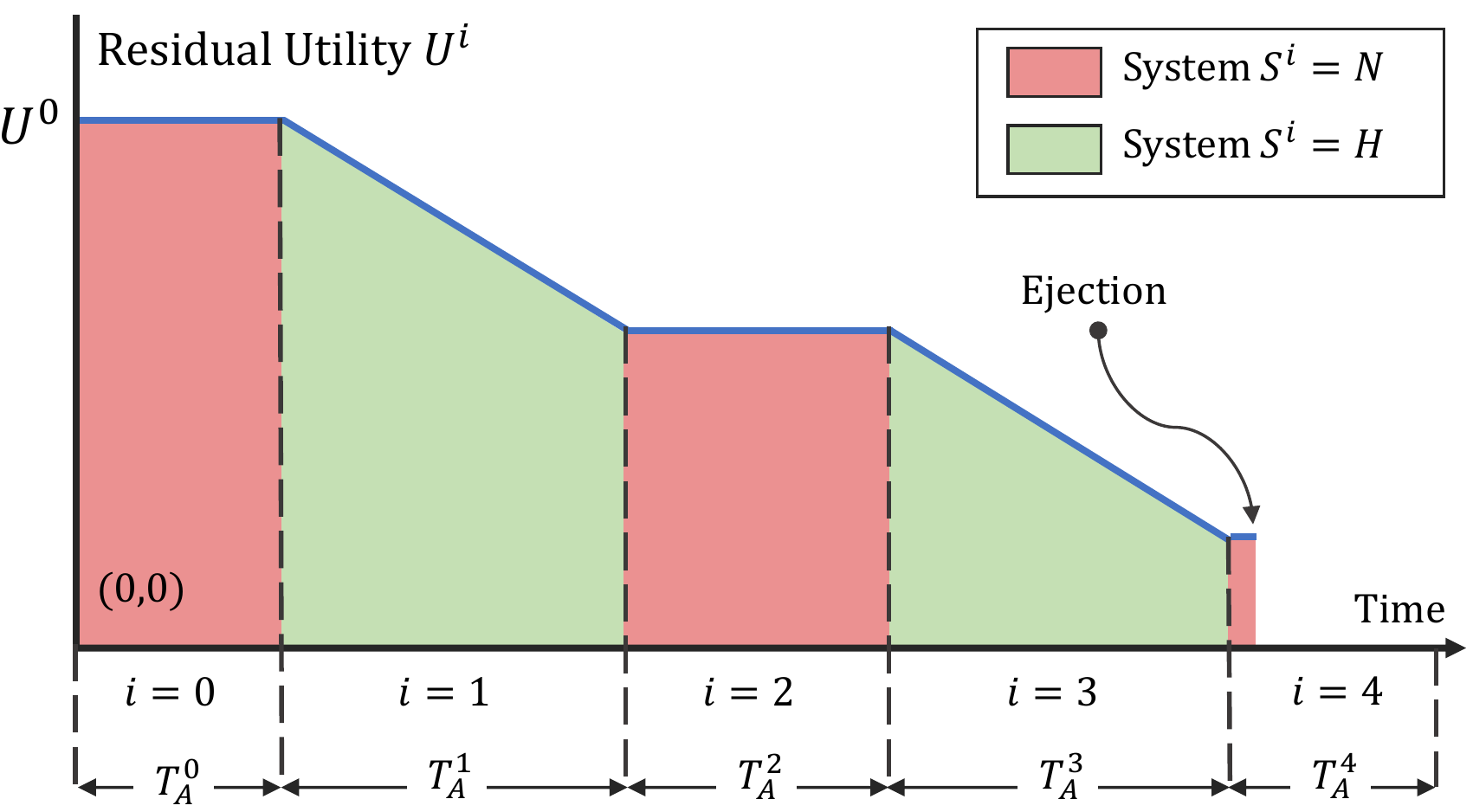} 
\par\end{centering}

\caption{\label{fig:latMove}$A$ moves throughout a network between honeypots
$H$ and normal systems $N.$ $D$ can earn a total of $U^{0}$ utility
for investigating $A.$ When $A$ is in a honeypot, $D$ learns and
the residual utility for future investigation decreases. Near $U^{i}=0,$
the risk of exposure outweighs the benefit of surveillance, and $D$
ejects $A$ at stage $i=4$ (in this example).}
\end{figure}
A discrete-time, continuous state MDP can be summarized by the tuple
$\left\langle \mathbb{X},\mathbb{A},\mu,q\right\rangle ,$ where $\mathbb{X}$
is the continuous state space, $\mathbb{A}$ is the set of actions,
$\mu:\,\mathbb{X}\times\mathbb{A}\to\mathbb{R}$ is the reward function,
and $q:\,\mathbb{X}\times\mathbb{A}\times\mathbb{X}\to\mathbb{R}_{+}$
is the transition kernel. In this section, we describe each of the
elements of $\left\langle \mathbb{X},\mathbb{A},\mu,q\right\rangle .$

\subsection{State Space $\mathbb{X}$\label{sub:State-Space}}

An attacker $A$ moves throughout a network containing two types of
systems $S$: honeypots $H$ and normal systems $N.$ At any time,
a network defender $D$ can eject $A$ from the network. $L$ denotes
having left the network. Together, we have $S\in\mathbb{S}\triangleq\{H,N,L\}.$ 

Let $i\in0,1,2,\ldots$ denote the discrete \emph{stage} of the game,
\emph{i.e.}, $i$ indicates the order of the systems visited\footnote{We consider a large network in which $A$ does not revisit individual
honeypots or normal systems, although he may visit multiple honeypots
and multiple normal systems.}. $D$ observes the types $S^{i}$ of the systems that $A$ visits.
The attacker, on the other hand, does not know the system types. 

We assume that there is a maximum amount of information that $D$
can learn from investigating $A.$ Let $U^{0}$ denote the corresponding
utility that $D$ receives for this information. At stage $i\in0,1,2,\ldots,$
let $U^{i}\in\mathbb{U}\triangleq[0,U^{0}]$ denote the \emph{residual
utility} available to $D$ for investigating $A.$ For instance, at
$i=5,$ $D$ may have recorded the attacker's time of infiltration,
malware type and operating system, but not yet any privilege escalation
attempts, which could reveal the attacker's objective. In that case,
$D$ may estimate that $U^{5}\approx0.6U^{0},$ \emph{i.e.}, $D$
has learned approximately $60\%$ of all possible information about
$A.$ 

$D$ should use $U^{i}$ together with $S^{i}$ to form his policy.
For instance, with $U^{5}\approx0.6U^{0},$ $D$ may allow $A$ to
remain in a honeypot $S^{5}=H$. But after observing a privilege escalation
attempt, with $U^{6}\approx0.8U^{0},$ $D$ may eject $A$ from $S^{6}=H,$
since there is little more to be learned about him. Therefore, $U^{i}$
and $S^{i}$ are both states. The full state space is $\mathbb{X}=\mathbb{U}\times\mathbb{S}.$
Figure \ref{fig:latMove} summarizes the interaction.

\subsection{One-Stage Actions $\mathbb{A}$\label{sub:Defender-Action}}

Let $\aleph_{0}$ denote the cardinality of the set of natural numbers
and $\mathbb{R}_{+}$ denote the set of non-negative real numbers.
Then define $T_{D}=\{T_{D}^{0},T_{D}^{1},T_{D}^{2},\ldots\}\in\mathbb{R}_{+}^{\aleph_{0}}$
such that $T_{D}^{i}$ denotes the time that $D$ plans to wait at
stage $i$ before ejecting $A$ from the network. The single-stage
action of $D$ is to choose $T_{D}^{i}\in\mathbb{A}=\mathbb{R}_{+}.$

\subsection{Reward Function $\mu$\label{sub:Reward-Function} }

To formulate the reward, we also need to define $T_{A}=\{T_{A}^{0},T_{A}^{1},T_{A}^{2},\ldots\}\in\mathbb{R}_{+}^{\aleph_{0}}.$
For each $i\in0,1,2,\ldots,$ $T_{A}^{i}$ denotes the duration of
time that $A$ plans to wait at stage $i$ before changing to a new
system\footnote{$D$ \emph{plans to wait,} because $A$ may move before $D$ ejects
him. Similarly, $A$ \emph{plans to wait}, because he may be ejected
from the network before this time has elapsed. But hereafter, we simply
say $A$ and $D$ \emph{wait}. }. Let $C_{N}<0$ denote the average cost per unit time that $D$ incurs
while $A$ resides in normal systems\footnote{Future work can consider different costs for each individual system
in a structured network.}. This cost may be estimated by a sum of the costs $\phi_{j}^{m}<0$
per unit time of each vulnerability $j\in1,2,\ldots,J$ on each the
systems $m\in1,2,\ldots,M$ in the network, weighted by the likelihoods
$\rho_{j}^{m}\in[0,1]$ that $A$ exploits the vulnerability: 
\[
C_{N}=\frac{1}{M}\stackrel[m=1]{M}{\sum}C_{N}^{m}=\frac{1}{M}\stackrel[m=1]{M}{\sum}\stackrel[j=1]{J}{\sum}\rho_{j}^{m}\phi_{j}^{m}.
\]
We also let $C_{H}\leq0$ denote a cost that $D$ pays to maintain
$A$ in a honeypot. This cost could represent, \emph{e.g.}, the expense
of hiring personnel to monitor the honeypot or the expense of redeployment.
Next, let $\mathbb{R}_{++}$ denote the set of strictly positive real
numbers. Let $v\in\mathbb{R}_{++}$ denote the utility per unit time
that $D$ gains from learning about $A$ while he is in honeypots. 

Define the function $\mu:\,\mathbb{U}\times\mathbb{S}\times\mathbb{R}_{+}\to\mathbb{R}$
such that $\mu(U^{i},S^{i},T_{D}^{i}\,|\,T_{A}^{i})$ gives the one-stage
reward to $D$ if the residual utility is $U^{i},$ $A$ is in system
$S^{i},$ $A$ waits for $T_{A}^{i}$ before moving, and $D$ waits
for $T_{D}^{i}$ before ejecting $A.$ Let $T^{i}\triangleq\min(T_{A}^{i},T_{D}^{i})$
denote the time for which $A$ remains at system $S^{i}$ before moving
or being ejected. Also let $\mathbf{1}\{P\}$ be the indicator function
which returns $1$ if it the statement $P$ is true. We have $\mu(U^{i},S^{i},T_{D}^{i}\,|\,T_{A}^{i})=$
\[
\mathbf{1}\{S=N\}C_{N}T^{i}+\mathbf{1}\{S=H\}\left(\min\left(T^{i}v,U^{i}\right)+C_{H}T^{i}\right).
\]

\subsection{Transition Kernel $q$\label{sub:Transition-Kernal}}

Let $\mathbb{R}_{+}$ denote the set of non-negative real numbers.
For stage $i\in0,1,2,\ldots,$ and given attacker and defender move
times $T_{A}^{i}$ and $T_{D}^{i},$ respectively, define the transition
kernel $q:\,\mathbb{U}\times\mathbb{S}\times\mathbb{R}_{+}\times\mathbb{U}\times\mathbb{S}\to\mathbb{R}_{+}$
such that, for all residual utilities $U^{i}\in\mathbb{U}$ and system
types $S^{i}\in\mathbb{S}$
\[
\int_{U^{i+1}\in\mathbb{U}}\int_{S^{i+1}\in\mathbb{S}}q\left(U^{i+1},S^{i+1},T_{D}^{i},U^{i},S^{i}\,|\,T_{A}^{i}\right)=1,
\]
where $U^{i+1}$ and $S^{i+1}$ denote the residual utility and system
type, respectively, at the next stage.

Let $p\in[0,1]$ denote the fraction of normal systems in the network\footnote{Again, in a formal network, the kernel will differ among different
honeypots and different normal systems. The fraction $p$ is an approximation
which is exact for a fully-connected network.}. For a real number $y,$ let $\delta(y)$ be the Dirac delta function.
For brevity, let $\Phi(U^{i},T)\triangleq\max\{U^{i}-vT,0\}.$ If
$T_{A}^{i}>T_{D}^{i},$ then $D$ ejects $A$ from the system, and
we have $q(U^{i+1},S^{i+1},T_{D}^{i},U^{i},S^{i}\,|\,T_{A}^{i})=$
\begin{multline}
\mathbf{1}\left\{ S^{i}=L\cap S^{i+1}=L\right\} \delta\left(U^{i+1}-U^{i}\right)+\\
\mathbf{1}\left\{ S^{i}=N\cap S^{i+1}=L\right\} \delta\left(U^{i+1}-U^{i}\right)+\\
\mathbf{1}\left\{ S^{i}=H\cap S^{i+1}=L\right\} \delta\left(U^{i+1}-\Phi(U^{i},T_{D}^{i})\right).\label{eq:transProbEject}
\end{multline}
If $T_{A}^{i}\leq T_{D}^{i},$ then $A$ changes systems, and we have
$q\left(U^{i+1},S^{i+1},T_{D}^{i},U^{i},S^{i}\,|\,T_{A}^{i}\right)=$
\begin{multline}
p\mathbf{1}\left\{ S^{i}=N\cap S^{i+1}=N\right\} \delta\left(U^{i+1}-U^{i}\right)+\\
\left(1-p\right)\mathbf{1}\left\{ S^{i}=N\cap S^{i+1}=H\right\} \delta\left(U^{i+1}-U^{i}\right)+\\
p\mathbf{1}\left\{ S^{i}=H\cap S^{i+1}=N\right\} \delta\left(U^{i+1}-\Phi(U^{i},T_{A}^{i})\right)+\\
\left(1-p\right)\mathbf{1}\left\{ S^{i}=H\cap S^{i+1}=H\right\} \delta\left(U^{i+1}-\Phi(U^{i},T_{A}^{i})\right).\label{eq:transProbCtu}
\end{multline}

\subsection{Infinite-Horizon, Undiscounted Reward\label{sub:Infinite-Horizon,-Undiscounted-R}}

For stage $i\in0,1,2,\ldots,$ define the stationary deterministic
feedback policy $\theta:\,\mathbb{U}\times\mathbb{S}\to\mathbb{R}_{+}$
such that $T_{D}^{i}=\theta(U^{i},S^{i})$ gives the time that $D$
waits before ejecting $A$ if the residual utility is $U^{i}$ and
the system type is $S^{i}.$ Let $\Theta$ denote the space of all
such stationary policies. Define the expected infinite-horizon, undiscounted
reward by $\mathcal{V}_{\theta}^{i}:\,\mathbb{U}\times\mathbb{S}\to\mathbb{R}$
such that $\mathcal{V}_{\theta}^{i}(U^{i},S^{i})$ gives the expected
reward from stage $i$ onward for using the policy $\theta$ when
the residual utility is $U^{i}$ and the type of the system is $S^{i}.$
We have 
\[
\mathcal{V}_{\theta}^{i}\left(U^{i},S^{i}\right)=\mathbb{E}\left\{ \stackrel[k=i]{\infty}{\sum}\mu\left(U^{k},S^{k},\theta\left(U^{k},S^{k}\right)\,|\,T_{A}^{k}\right)\right\} ,
\]
such that the states transition according to Eq. (\ref{eq:transProbEject}-\ref{eq:transProbCtu}).
Given an initial system type $S^{0}\in\{H,N\},$ the overall problem
for $D$ is to find $\theta^{*}$ such that  
\[
\theta^{*}\in\underset{\theta\in\Theta}{\arg\max}\,\mathcal{V}_{\theta}^{0}\left(U^{0},S^{0}\right).
\]
The undiscounted utility function demands Proposition \ref{prop:expValFinite}.
\begin{prop}
\label{prop:expValFinite}$\mathcal{V}_{\theta^{*}}^{i}(U^{i},S^{i})$
is finite.\end{prop}
\begin{IEEEproof}
See Appendix \ref{sec:proofExpValFinite}. 
\end{IEEEproof}
It is also convenient to define the \emph{value function} as the reward
for the optimal policy: 
\[
\mathcal{V}^{i}\left(U^{i},S^{i}\right)\triangleq\mathcal{V}_{\theta^{*}}^{i}\left(U^{i},S^{i}\right)=\underset{\theta\in\Theta}{\max}\,\mathcal{V}_{\theta}^{i}\left(U^{i},S^{i}\right).
\]
The Bellman principle \cite{bellman1952theory} implies that for an
optimal stationary policy $\theta^{*},$ and for $i\in0,1,2,\ldots,$
$\theta^{*}\left(U^{i},S^{i}\right)\in$ 
\begin{multline*}
\underset{T_{D}^{i}\in\mathbb{R}_{+}}{\arg\max}\,\mu\left(U^{i},S^{i},T_{D}^{i}\,|\,T_{A}^{i}\right)+\int_{U^{i+1}\in\mathbb{U}}\int_{S^{i+1}\in\mathbb{S}}\\
\mathcal{V}^{i+1}\left(U^{i+1},S^{i+1}\right)q(U^{i+1},S^{i+1},T_{D}^{i},U^{i},S^{i}\,|\,T_{A}^{i}).
\end{multline*}

\section{Analysis and Results}

\label{sec:anaDef}
\begin{figure*}
\subfloat[\label{fig:val6}$p=0.60,$ $\omega\approx0.83,$ $\delta=3.0$]{\centering{}\includegraphics[width=1\columnwidth]{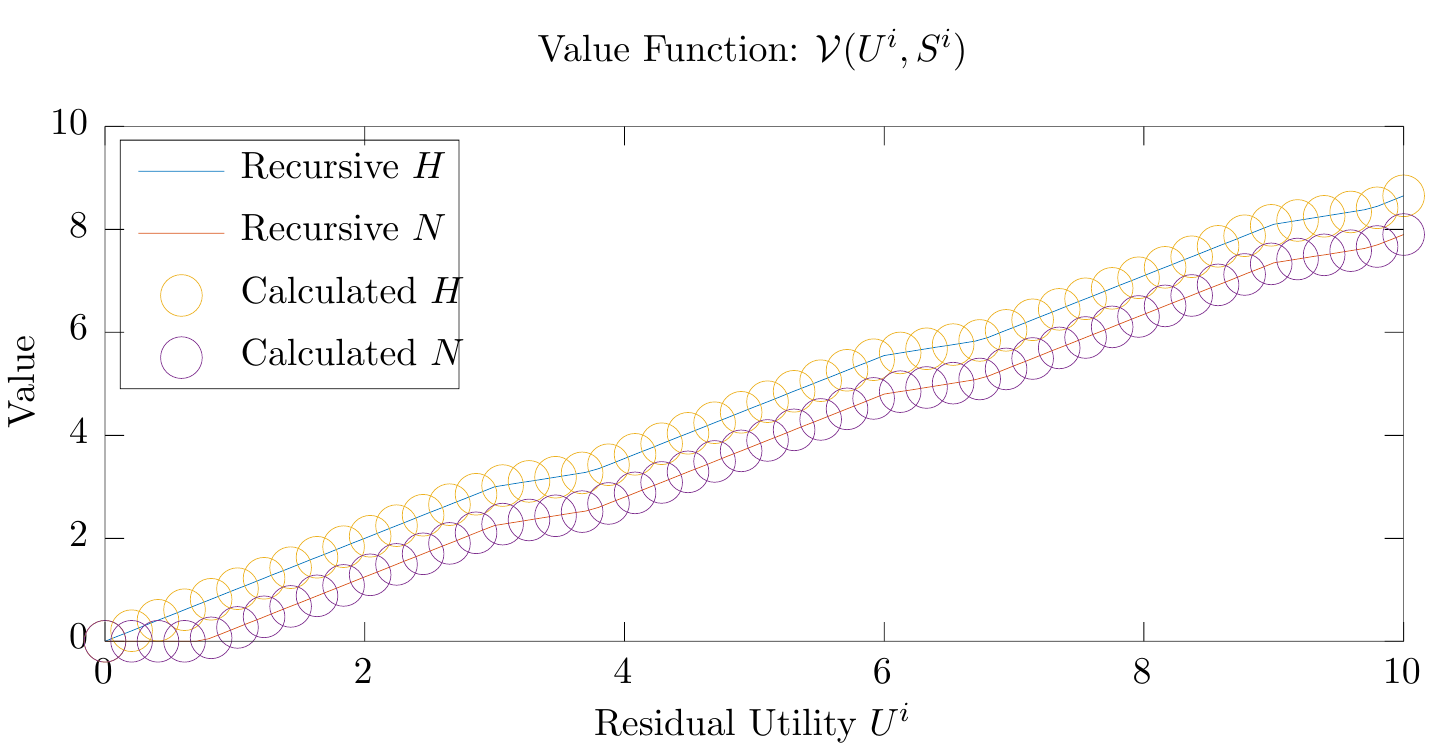} }\hfill{}\subfloat[\label{fig:val8}$p=0.85,$ $\omega\approx2.2,$ $\delta=3.0$]{\centering{}\includegraphics[width=1\columnwidth]{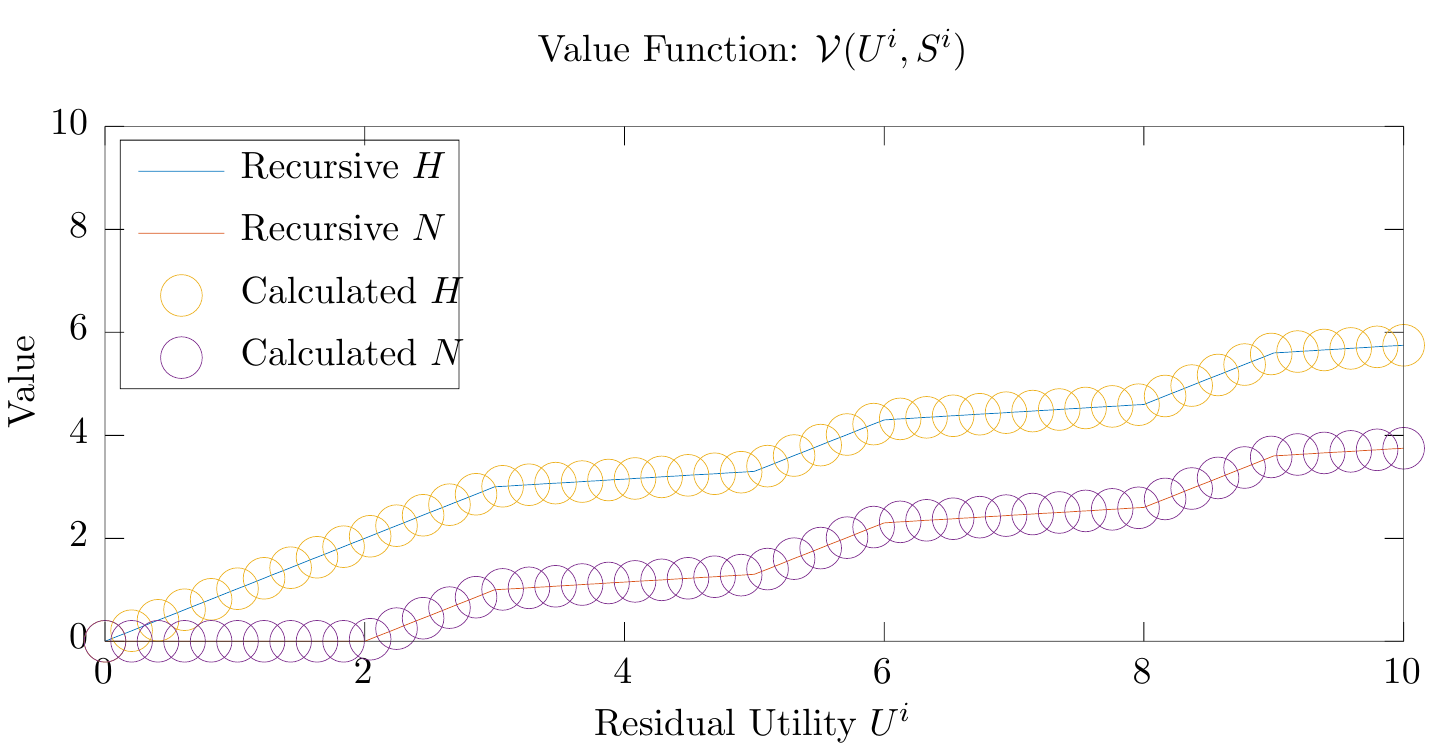} }

\caption{\label{fig:allValFns}Value functions with $p=0.60$ and $p=0.85.$
The top and bottom curves depict $\mathcal{V}(U^{i},H)$ and $\mathcal{V}(U^{i},N),$
respectively, as a function of $U^{i}.$ The circles plot the analytical
$\mathcal{V}(U^{i},S),$ $S\in\{H,N\}$ from Theorem \ref{thm:valueV},
and the solid lines verify this using an iterative numerical method.}
\end{figure*}
In this section, we solve for the value function and optimal policy.
We start by obtaining the optimal policy in honeypots, and reducing
the space of candidates for an optimal policy in normal systems. Then
we present the value function and optimal policy separately, although
they are derived simultaneously.

\subsection{Reduced Action Spaces}

Lemma \ref{lem:reducedHspace} obtains the optimal waiting time for
$S^{i}=H.$
\begin{lem}
\label{lem:reducedHspace}(Optimal Policy for $S^{i}=H$) In honeypots,
for any $i\in0,1,2,\ldots$ and $U^{i}\in\mathbb{U},$ the value function
is optimized by playing $T_{D}^{i}=U^{i}/v.$\end{lem}
\begin{IEEEproof}
The value of the game is maximized if $A$ passes through only honeypots
and $D$ ejects $A$ when the residual utility is $0.$ $D$ can achieve
this by playing $T_{D}^{i}=U^{i}/v$ if $T_{A}^{i}>U^{i}/v.$ On the
other hand, if $T_{A}^{i}\leq U^{i}/v,$ then it is optimal for $D$
to allow $A$ to change systems. This is optimal because the value
function at stage $i+1$ is non-negative, since in the worst case
$D$ can eject $A$ immediately if $A$ arrives at a normal system.
$D$ can allow $A$ to change systems by playing any $T_{D}^{i}\geq T_{A}^{i},$
although it is convenient for brevity of notation to choose $T_{D}^{i}=T_{A}^{i}.$ 
\end{IEEEproof}
Lemma \ref{lem:reducedNspace} narrows the optimal waiting times for
$S^{i}=N.$
\begin{lem}
\label{lem:reducedNspace}(Reduced Action Space for $S^{i}=N$) In
normal systems, for any $i\in0,1,2,\ldots$ and $U^{i}\in\mathbb{U},$
the value function is optimized by playing either $T_{D}^{i}=0$ or
$T_{D}^{i}=T_{A}^{i}.$ \end{lem}
\begin{IEEEproof}
First, note that it is always suboptimal for $D$ to eject $A$ at
a time less that $T_{A}^{i}.$ That is, for stage $i\in0,1,2,\ldots,$
$\mathcal{V}_{\tilde{\theta}}^{i}(U^{i},N)<\mathcal{V}_{\hat{\theta}}^{i}(U^{i},N)$
for $0=\hat{\theta}(U^{i},N)<\tilde{\theta}(U^{i},N)<T_{A}^{i}.$
Second, note that $D$ receives the same utility for ejecting $A$
at any time greater than or equal to $T_{A}^{i},$ \emph{i.e.}, $\mathcal{V}_{\tilde{\theta}}^{i}(U^{i},N)=\mathcal{V}_{\hat{\theta}}^{i}(U^{i},N)$
for $T_{A}^{i}\leq\hat{\theta}(U^{i},N)\leq\tilde{\theta}(U^{i},N).$
Then either $0$ or $T_{A}^{i}$ is optimal.
\end{IEEEproof}
Remark \ref{rem:reducedSpaces} summarizes Lemmas \ref{lem:reducedHspace}-\ref{lem:reducedNspace}.
\begin{rem}
\label{rem:reducedSpaces} Lemma \ref{lem:reducedHspace} obtains
the unique optimal waiting time in honeypots. Lemma \ref{lem:reducedNspace}
reduces the candidate set of optimal waiting times in normal systems
to two times: $T_{D}^{i}\in\{0,T_{A}^{i}\}.$ These times are equivalent
to stopping the Markov chain and allowing it to continue, respectively.
Thus, Lemmas \ref{lem:reducedHspace}-\ref{lem:reducedNspace} show
that the MDP is an optimal stopping problem.
\end{rem}

\subsection{Value Function Structure}

To solve the optimal stopping problem, we must find the value function.
We obtain the value function for a constant attacker action, \emph{i.e.},
$T_{A}^{0}=T_{A}^{1}=....\triangleq\bar{T}_{A}.$ This means that
$\mathcal{V}^{i}\equiv\mathcal{V}.$ Define the following notation:
\begin{equation}
\delta\triangleq\bar{T}_{A}v,\;\;\delta_{1}^{D}\triangleq\bar{T}_{A}\left(v+C_{H}\right),\label{eq:notationD1}
\end{equation}
\begin{equation}
\lambda_{N}^{D}\triangleq\frac{-C_{N}}{1-p},\;\;\chi_{H}^{D}\triangleq\frac{v+C_{H}}{v}.\label{eq:notationD2}
\end{equation}
Note that $\delta$ and $\delta_{1}^{D}$ are in units of utility,
$\lambda_{N}^{D}$ is in units of utility per second, and $\chi_{H}^{D}$
is unitless. 

First, $\mathcal{V}(U^{i},L)=0$ for all $U^{i}\in\mathbb{U},$ because
no further utility can be earned after $D$ ejects $A.$ Next, $\mathcal{V}(0,S)=0$
for both $S\in\{H,N\},$ because no positive utility can be earned
in either type of system. $\mathcal{V}$ can now be solved backwards
in $U^{i}$ from $U^{i}=0$ to $U^{i}=U^{0}$ using these terminal
conditions. Depending on the parameters, it is possible that $\forall U^{i}\in\mathbb{U},$
$\theta^{*}(U^{i},N)=0$ and $\mathcal{V}(U^{i},N)=0,$ \emph{i.e.},
$D$ should eject $A$ from all normal systems immediately. We call
this the trivial case. Lemma \ref{lem:omegaExists} describes the
structure of the optimal policy outside of the trivial case.
\begin{lem}
\label{lem:omegaExists}(Optimal Policy Structure) Outside of the
trivial case, there exists a residual utility $\omega\in\mathbb{U}$
such that:\end{lem}
\begin{itemize}
\item for $U^{i}<\omega,$ $\theta^{*}(U^{i},N)=0$ and $\mathcal{V}(U^{i},N)=0,$
\item for $U^{i}>\omega,$ $\theta^{*}(U^{i},N)=\bar{T}_{A}$ and $\mathcal{V}(U^{i},N)>0.$\end{itemize}
\begin{IEEEproof}
See Appendix \ref{sec:proofV}.
\end{IEEEproof}

\subsection{Value Function Threshold}

Next, for $x\in\mathbb{R},$ define
\begin{equation}
k\left[x\right]\triangleq\begin{cases}
\left\lfloor x/\delta\right\rfloor , & \text{if }x\geq0\\
0, & \text{if }x<0
\end{cases},\label{eq:defK}
\end{equation}
where $\left\lfloor \bullet\right\rfloor $ is the floor function.
The floor function is required because $\mu$ is nonlinear in $U^{i}.$
Then Theorem \ref{thm:omegaClosedForm} gives $\omega$ in closed
form.
\begin{thm}
\label{thm:omegaClosedForm}(Threshold $\omega$) Outside of the trivial
case, the threshold $\omega$ of residual utility beyond which $D$
should eject $A$ is given by
\[
\omega=\delta\left(k\left[\omega\right]+\frac{\lambda_{N}^{D}}{\left(v+C_{H}\right)\left(1-p\right)^{k\left[\omega\right]}}-\frac{1-\left(1-p\right)^{k\left[\omega\right]}}{p\left(1-p\right)^{k\left[\omega\right]}}\right),
\]
where $k[\omega]$ is defined as in Eq. (\ref{eq:defK}), and it
can be shown that
\[
k\left[\omega\right]=\left\lfloor \log_{1-p}\left(1+\frac{pC_{N}}{\left(1-p\right)\left(v+C_{H}\right)}\right)\right\rfloor ,
\]
if the argument of the logarithm is positive. If not, then the optimal
policy is for $D$ to eject $A$ from normal systems immediately.\end{thm}
\begin{IEEEproof}
See Appendix \ref{sec:proofOmega}.
\end{IEEEproof}
Remark \ref{rem:intuitionOmega} gives some intuition about Theorem
\ref{thm:omegaClosedForm}.
\begin{rem}
\label{rem:intuitionOmega}Numerical results suggest that in many
cases (such as those in Fig. \ref{fig:allValFns}), $k[\omega]=0.$
In that case, we have $\omega=-\delta C_{N}/\left((v+C_{H})(1-p)\right).$
The threshold $\omega$ increases as the cost for normal systems ($C_{N}$)
increases, decreases as the rate at which utility is gained in normal
systems ($v$) increases, and decreases as the proportion of normal
systems ($p$) increases.
\end{rem}
Finally, Theorem \ref{thm:valueV} summarizes the value function.
\begin{thm}
\label{thm:valueV}(Value Function) The value function is given by
\[
\mathcal{V}\left(U^{i},S^{i}\right)=\begin{cases}
0, & \text{if }S^{i}=L\\
f^{D}(U^{i}), & \text{if }S^{i}=H\\
\left\{ f^{D}(U^{i})-\bar{T}_{A}\lambda_{N}^{D}\right\} _{+}, & \text{if }S^{i}=N
\end{cases},
\]
where $\{\bullet\}_{+}$ denotes $\max\{\bullet,0\},$ and $f^{D}:\,\mathbb{U}\to\mathbb{R}_{+}$
is
\begin{multline*}
f^{D}\left(U^{i}\right)\triangleq\chi_{H}^{D}\left(U^{i}-\delta k[U^{i}]\right)\left(1-p\right)^{k[U^{i}]-k[U^{i}-\omega]}+\\
\frac{\delta_{1}^{D}}{p}\left(1-(1-p)^{k[U^{i}]-k[U^{i}-\omega]}\right)+k[U^{i}-\omega]\left(\delta_{1}^{D}-p\lambda_{N}^{D}\bar{T}_{A}\right).
\end{multline*}
\end{thm}
\begin{IEEEproof}
See Appendix \ref{sec:proofV}. 
\end{IEEEproof}
Remark \ref{rem:valueFn} discusses the interpretation of Theorem
\ref{thm:valueV}.
\begin{rem}
\label{rem:valueFn}The quantity $f^{D}(U^{i})$ is the expected reward
for future surveillance, while $\bar{T}_{A}\lambda_{N}^{D}$ is the
expected damage that will be caused by $A.$ In normal systems, when
$U^{i}\leq\omega,$ we have $f^{D}(U^{i})\leq\bar{T}_{A}\lambda_{N}^{D},$
and the risk of damage outweighs the reward of future surveillance.
Therefore, it is optimal for $D$ to eject $A,$ and $\mathcal{V}(U^{i},N)=0.$
On the other hand, for $U^{i}>\omega,$ it is optimal for $D$ to
allow $A$ to remain for $\bar{T}_{A}$ before moving, so $\mathcal{V}(U^{i},N)>0.$
Figure \ref{fig:allValFns} gives examples of the value function.
\end{rem}

\subsection{Optimal Policy Function}

Theorem \ref{thm:optDef} summarizes the optimal policy.
\begin{thm}
\label{thm:optDef}(Defender Optimal Policy) $D$ achieves an optimal
policy for $S^{i}\in\{H,N\}$ by playing 

\[
\theta^{*}\left(U^{i},S^{i}\right)=\begin{cases}
U^{i}/v, & \text{if }S^{i}=H\\
\bar{T}_{A}, & \text{if }S^{i}=N\text{ and }U^{i}\geq\omega\\
0, & \text{if }S^{i}=N\text{ and }U^{i}<\omega
\end{cases}.
\]
\end{thm}
\begin{IEEEproof}
See Appendix \ref{sec:proofV}. 
\end{IEEEproof}
Remark \ref{rem:onlyU} gives an observation about the optimal policy.
\begin{rem}
\label{rem:onlyU}During attacker engagement, Theorem \ref{thm:optDef}
only requires estimating $U^{i},$ \emph{i.e.}, the remaining information
which can be learned about the attacker. $D$ will allow $A$ to remain
in the network until $U^{i}<\omega.$ The cumulative information lost
in stages $k\in0,1,\ldots,i$ need not be known, since it is not part
of the state.\end{rem}

\section{Robustness Evaluation}

\label{anaAtk}

In this section, we evaluate the robustness of the policy $\theta^{*}$
by allowing $A$ to choose the worst-case $\bar{T}_{A}.$

\subsection{Equilibrium Concept\label{sub:Equilibrium-Concept}}

Let us write $\mathcal{V}_{\theta}(U^{i},S^{i}\,|\,\bar{T}_{A})$
and $\theta^{*}(U^{i},S^{i},\,|\,\bar{T}_{A})$ to denote the dependence
of the value and optimal policy, respectively, on $\bar{T}_{A}.$
Next, define $\bar{\mathcal{V}}:\,\mathbb{R}_{+}\to\mathbb{R}$ such
that $\bar{\mathcal{V}}(\bar{T}_{A})$ gives the expected utility
to $D$ over possible types of initial systems for playing $\theta^{*}$
as a function of $\bar{T}_{A}.$ This is given by
\begin{equation}
\bar{\mathcal{V}}\left(\bar{T}_{A}\right)=p\mathcal{V}\left(U^{0},N\,|\,\bar{T}_{A}\right)+\left(1-p\right)\mathcal{V}\left(U^{0},H\,|\,\bar{T}_{A}\right).\label{eq:totDefUtil}
\end{equation}

Definition \ref{def:stackEq} formulates a \emph{zero-sum} \emph{Stackelberg
equilibrium }\cite{vonstackelbergmarktform1934} in which $A$ chooses
$\bar{T}_{A}$ to minimize Eq. (\ref{eq:totDefUtil}), and $D$ plays
the optimal policy given $\bar{T}_{A}$ from Theorem \ref{thm:optDef}.
\begin{defn}
(Stackelberg Equilibrium)\label{def:stackEq} A Stackelberg equilibrium
(SE) of the zero-sum attacker-defender game is a strategy pair $(\bar{T}_{A}^{*},\theta^{*})$
such that 
\[
\bar{T}_{A}^{*}\in\underset{\bar{T}_{A}}{\arg\min}\:\bar{\mathcal{V}}_{\theta^{*}(U^{i},S^{i}\,|\,\bar{T}_{A})}\left(\bar{T}_{A}\right),
\]
and $\forall U^{i}\in\mathbb{U},$ $\forall S^{i}\in\mathbb{S},$
\[
\theta^{*}\left(U^{i},S^{i}\,|\,\bar{T}_{A}^{*}\right)\in\underset{\theta\in\Theta}{\arg\max}\,\mathcal{V}_{\theta}\left(U^{i},S^{i}\,|\,\bar{T}_{A}^{*}\right).
\]

\end{defn}
Definition \ref{def:stackEq} considers $A$ as the Stackelberg game
leader because our problem models an intelligent defender who reacts
to the strategy of an observed attacker.

\subsection{Equilibrium Analysis\label{sub:Equilibrium-Analysis}}

\begin{figure}
\begin{centering}
\includegraphics[width=0.8\columnwidth]{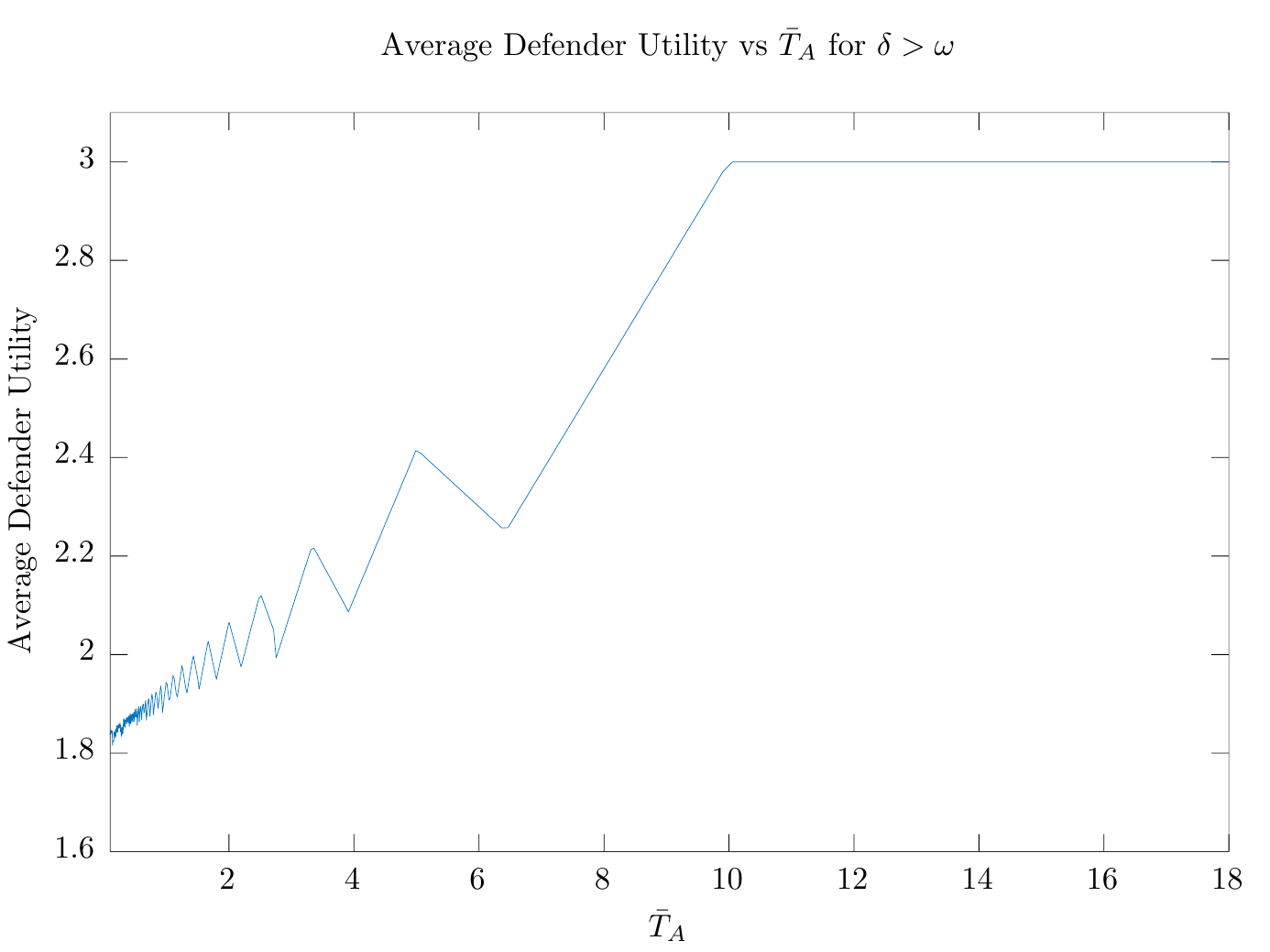} 
\par\end{centering}

\caption{\label{fig:Vta_delLessOm}$\bar{\mathcal{V}}_{\theta^{*}}(\bar{T}_{A})$
for the case that $\delta<\omega.$ Here, the worst case value is
$\bar{\mathcal{V}}_{\theta^{*}}(\bar{T}_{A}^{*})\approx1.8,$ which
occurs as $\bar{T}_{A}\to0.$ }
\end{figure}

\begin{figure}
\begin{centering}
\includegraphics[width=0.8\columnwidth]{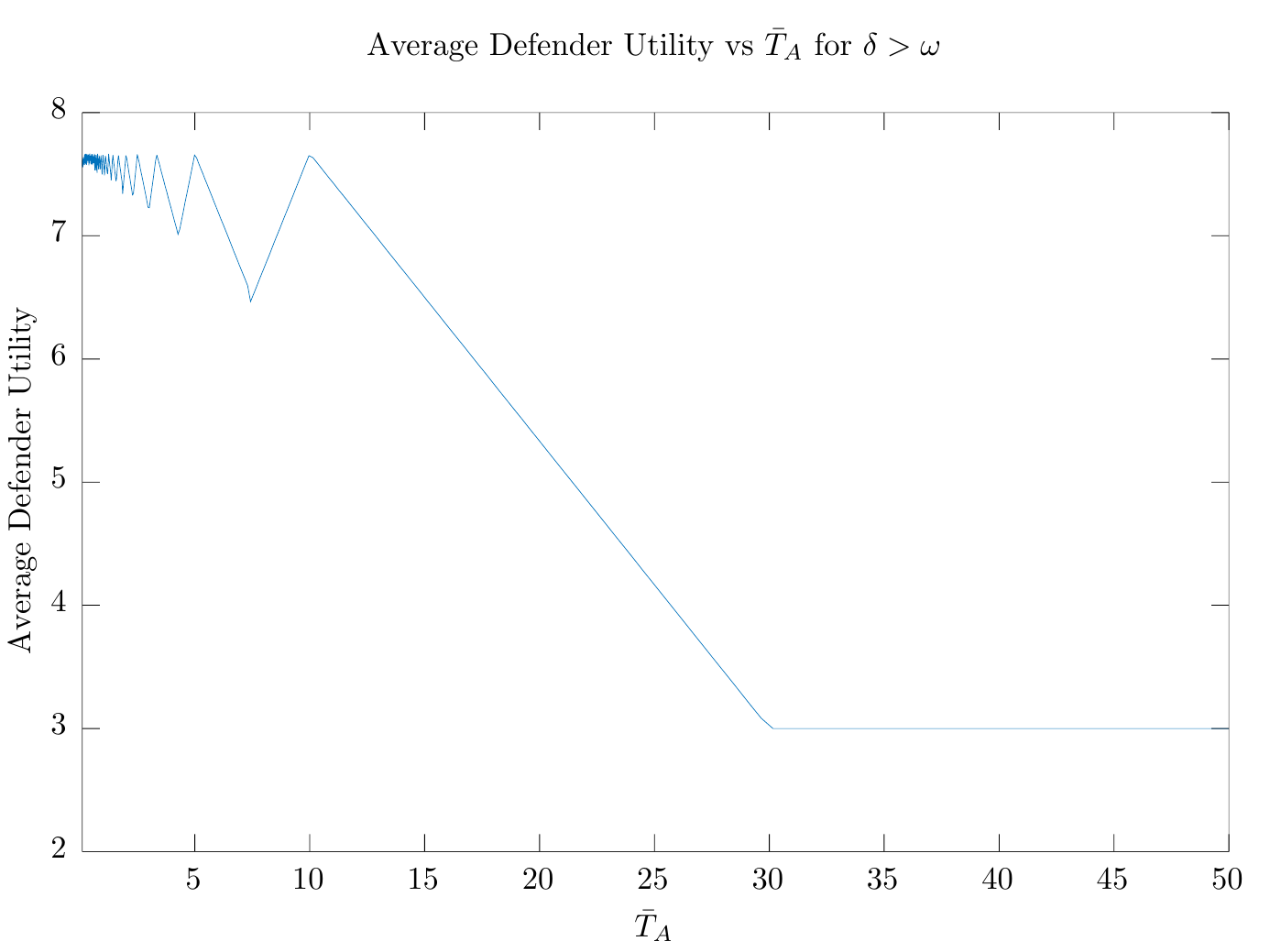} 
\par\end{centering}

\caption{\label{fig:Vta_delGrOm}$\bar{\mathcal{V}}_{\theta^{*}}(\bar{T}_{A})$
for the case that $\delta>\omega.$ Here, the worst case value is
$\bar{\mathcal{V}}_{\theta^{*}}(\bar{T}_{A}^{*})\approx3.0,$ which
occurs for $\bar{T}_{A}>\omega\approx30.$}
\end{figure}

$\bar{\mathcal{V}}_{\theta^{*}}(\bar{T}_{A})$ takes two possible
forms, based on the values of $\delta$ and $\omega$. Figure \ref{fig:Vta_delLessOm}
depicts $\bar{\mathcal{V}}_{\theta^{*}}(\bar{T}_{A})$ for $\delta<\omega,$
and Fig. \ref{fig:Vta_delGrOm} depicts $\bar{\mathcal{V}}_{\theta^{*}}(\bar{T}_{A})$
for $\delta>\omega.$ Note that the oscillations are not produced
by numerical approximation, but rather by the nonlinear value function.
The worst-case $\bar{T}_{A}^{*}$ is as small as possible for $\delta<\omega$
and is large for $\delta>\omega.$ Theorem \ref{thm:worstTaComb}
states this result formally. 

\begin{thm}
\label{thm:worstTaComb}(Value as a function of $\bar{T}_{A}$) For
low $\bar{T}_{A},$ we have
\begin{equation}
\underset{\bar{T}_{A}\to0}{\lim}\:\bar{\mathcal{V}}_{\theta^{*}}(\bar{T}_{A})=U^{0}\left(1+\frac{1}{v}\left(C_{H}+C_{N}\frac{p}{1-p}\right)\right).\label{eq:valLowComb}
\end{equation}
Define $\bar{T}_{\omega}$ as $\bar{T}_{A}$ such that $U^{0}=\omega.$
Then for $\bar{T}_{A}\geq\max\{r,\bar{T}_{\omega}\},$ we have 
\begin{equation}
\bar{\mathcal{V}}_{\theta^{*}}(\bar{T}_{A})=U^{0}\left(1-p\right)\frac{v+C_{H}}{v}.\label{eq:valHighComb}
\end{equation}
\end{thm}
\begin{IEEEproof}
See Appendix \ref{sec:proofWorstVal}. 
\end{IEEEproof}
Remarks \ref{rem:similarityRobust}-\ref{rem:dragRobust} discuss
Theorem \ref{thm:worstTaComb} and Fig. \ref{fig:Vta_delLessOm}-\ref{fig:Vta_delGrOm}.
\begin{rem}
\label{rem:similarityRobust}The parameters of Fig. \ref{fig:Vta_delLessOm}
and Fig. \ref{fig:Vta_delGrOm} differ only in $C_{N},$ which has
a higher absolute value in Fig. \ref{fig:Vta_delLessOm}. Since $C_{N}$
only affects $\bar{\mathcal{V}}_{\theta^{*}}(\bar{T}_{A})$ as $\bar{T}_{A}\to0,$
the plots are the same for high $\bar{T}_{A}.$ 
\end{rem}

\begin{rem}
\label{rem:dragRobust}The connection between Fig. \ref{fig:Vta_delLessOm}
and Fig. \ref{fig:Vta_delGrOm} can be visualized by translating the
left sides of the curves vertically, while the right sides remain
fixed (at $\bar{\mathcal{V}}_{\theta^{*}}(\bar{T}_{A})\approx3.0$).
This gives network designers an intuition of how the worst-case value
can be manipulated by changing the parameters of the game.
\end{rem}
Finally, Corollary \ref{cor:worstCaseValue} summarizes the worst-case
value.
\begin{cor}
(Worst-Case Value)\label{cor:worstCaseValue} The worst case value
$\bar{\mathcal{V}}_{\theta^{*}}(\bar{T}_{A}^{*})$ is approximated
by\footnote{We say approximated because it has not been proven that the oscillations
as $\bar{T}_{A}\to0$ exclude a transient below $U^{0}\left(1+\frac{1}{v}\left(C_{H}+C_{N}\frac{p}{1-p}\right)\right)$
for $\delta<\omega$ or $U^{0}\left(1-p\right)\frac{v+C_{H}}{v}$
for $\delta>\omega.$ }
\[
U^{0}\underset{\bar{T}_{A}}{\min}\:\left\{ \left(1+\frac{1}{v}\left(C_{H}+C_{N}\frac{p}{1-p}\right)\right),\left(1-p\right)\frac{v+C_{H}}{v}\right\} .
\]
\end{cor}

\section{Simulation}

In this section, we simulate a network which sustains five attacks
and implements $D$'s optimal policy $\theta^{*}.$ Consider the example
network depicted in Fig. \ref{fig:honeyNetwork} in Section \ref{sec:intro}.
This network has $16$ production nodes, including routers, wireless
access points, wired admin access, and a database. It also has sensors,
actuators, and controllers, which form part of a SCADA system. The
network has $4$ honeypots (in the top-right of the figure), configured
to appear as additional SCADA system components.

\begin{figure}
\centering{}\includegraphics[width=0.85\columnwidth]{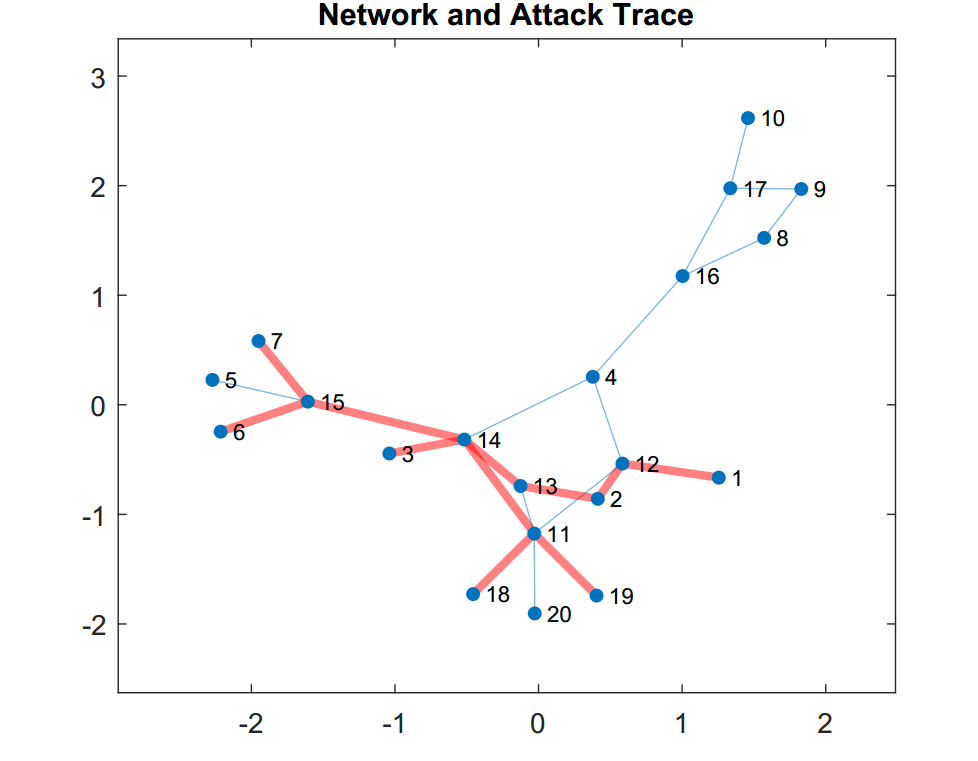}\caption{\label{fig:simNetwork}The blue nodes and edges illustrate a $20$-node
network, and the red highlights indicate an example attack trace.}
\end{figure}

Figure \ref{fig:simNetwork} depicts a view of the network in MATLAB
\cite{MATLAB2017}. The red line indicates an example attack path,
which enters through the wireless access point at node $1,$ passes
through the honeynet in nodes $11,$ $18,$ and $19,$ and enters
the SCADA components in nodes $6$ and $7.$ The transitions are realized
randomly.

\begin{figure}
\centering{}\includegraphics[width=0.85\columnwidth]{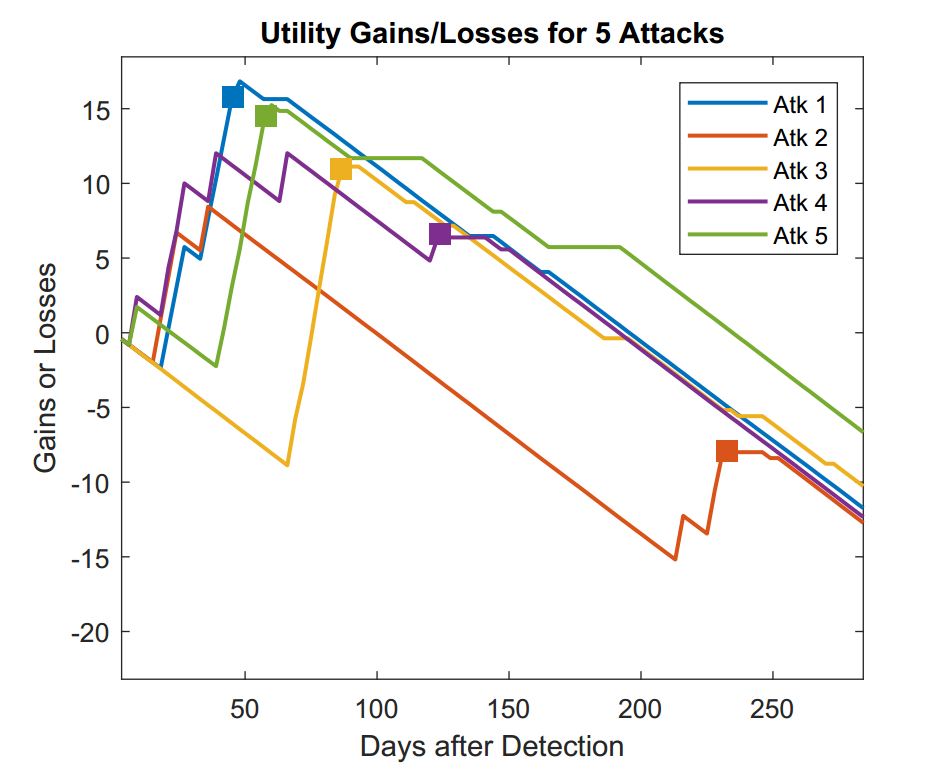}\caption{\label{fig:cumUtil}The curves indicate the cumulative utility gains
or losses for five simulated attacks. The solid squares indicate the
optimal ejection time according to $\theta^{*}.$}
\end{figure}

Figure \ref{fig:cumUtil} depicts the cumulative utility of $D$ over
time for five simulated attacks. Towards the beginning of the attacks,
$D$ gains utility. But after learning nears completion (\emph{i.e.},
$U^{i}\approx0$), the losses $C_{N}$ from normal systems dominate.
The filled boxes in each trace indicate the ejection point dictated
by $\theta^{*}.$ At these points, $U^{i}\leq\omega.$ The ejection
points are approximately at the maximum utility for traces $1,$ $3,$
and $5,$ and obtain a positive utility in trace $4.$ Trace $5$
involves a long period in which $S^{i}=N,$ and $D$ sustains heavy
losses. Since the traces are realized randomly, $\theta^{*}$ maximizes
expected utility rather than realized utility.

\section{Discussion of Results}

\label{sec:concl}

This paper aimed to assess how long an intelligent network defender
that detects an attacker should observe the attacker before ejecting
him. We found that the defender should keep the attacker in a honeypot
as long as information remains to be learned and in a normal system
until a threshold amount of information remains. This threshold is
$\omega,$ at which the benefits of observation exactly balance the
risks of information loss. Using this model, network designers can
vary parameters (\emph{e.g.}, the number of honeypots and the rate
at which they gather information) in order to maximize the value function
$\mathcal{V}.$ In particular, we have examined the effect of the
attacker move period $\bar{T}_{A}$ using a Stackelberg game in which
$A$ chooses the worst-case $\bar{T}_{A}.$ Future work can use signaling
games to calculate attacker beliefs $p$ and $1-p$ based on defender
strategies. Another direction, for distributed sensor-actuator networks,
is to quantify the risk $C_{N}$ of system compromise using optimal
control theory.

\appendices

\section{Proof of Finite Expected Value}

\label{sec:proofExpValFinite}

The maximum value of $\mathcal{V}_{\theta}^{i}(U^{i},S^{i})$ is achieved
if $A$ only visits honeypots. In this case, $\mathcal{V}_{\theta}^{i}(U^{i},S^{i})=(v+C_{H})U^{0}/v,$
so the expected utility is bounded from above. If $D$ chooses a poor
policy (for example, $\theta(U^{i},S^{i})=T_{A}^{i}$ for all $U^{i}\in\mathbb{U}$
and $S^{i}\in\mathbb{S}$), then $\mathcal{V}_{\theta}^{i}(U^{i},S^{i})$
can be unbounded below. On the other hand, $D$ can always guarantee
$\mathcal{V}_{\theta}^{i}(U^{i},S^{i})=0$ (for example, by choosing
$\theta(U^{i},S^{i})=0$ for all $U^{i}\in\mathbb{U}$ and $S^{i}\in\mathbb{S}$).
Therefore, the value of the \emph{optimal }policy is  bounded from
below as well as from above.

\section{Derivation of Value Function and Optimal Policy}

\label{sec:proofV}

For $S^{i}\in\{H,N\},$ the value function $\mathcal{V}(U^{i},S^{i})$
is piecewise-linear in $U^{i}.$ Let $\mathcal{V}(U^{i},S^{i})[a,b]$
denote $\mathcal{V}(U^{i},S^{i})$ restricted to the domain $U^{i}\in[a,b]\subset\mathbb{R}.$
First, we find $\mathcal{V}(U^{i},N)$ in terms of $\mathcal{V}(U^{i},H).$
For any non-negative integer $k,$ one step of the Bellman equation
gives $\mathcal{V}(U^{i},N)[k\delta,\left(k+1\right)\delta]=$ 
\begin{multline*}
\left\{ C_{N}\bar{T}_{A}+p\mathcal{V}\left(U^{i},N\right)\left[k\delta,\left(k+1\right)\delta\right]\right.\\
\left.\left.+\left(1-p\right)\mathcal{V}\left(U^{i},H\right)\left[k\delta,\left(k+1\right)\delta\right]\right]\right\} _{+},
\end{multline*}
where $\left\{ \bullet\right\} _{+}$ denotes $\max\{\bullet,0\}.$
$D$ achieves this maximization by continuing the game if the expected
value for continuing is positive, and ejecting $A$ if the expected
value is negative.

Rearranging terms and using Eq. (\ref{eq:notationD1}-\ref{eq:notationD2})
gives $\mathcal{V}(U^{i},N)\left[k\delta,\left(k+1\right)\delta\right]=\left\{ \mathcal{V}\left(U^{i},H\right)\left[k\delta,\left(k+1\right)\delta\right]-\lambda_{N}^{D}\bar{T}_{A}\right\} _{+}.$
Now, we have defined $\omega$ as $U^{i}\in\mathbb{R}_{+}$ which
makes the argument on the right side equal to zero. This obtains $\mathcal{V}\left(U^{i},N\right)\left[k\delta,\left(k+1\right)\delta\right]=$
\[
\begin{cases}
0, & \text{if }U^{i}\leq\omega\\
\mathcal{V}\left(U^{i},H\right)\left[k\delta,\left(k+1\right)\delta\right]-\lambda_{N}^{D}\bar{T}_{A}, & \text{if }U^{i}>\omega.
\end{cases}
\]

Next, we find $\mathcal{V}(U^{i},H).$ First, consider $\mathcal{V}(U^{i},H)[0,\delta].$
$D$ keeps $A$ in the honeypot until all residual utility is depleted,
and then ejects him. Thus $\mathcal{V}(U^{i},H)[0,\delta]=U^{i}\chi_{H}^{D}.$
Next, for $k\in1,2,\ldots,$ consider $\mathcal{V}(U^{i},H)[k\delta,(k+1)\delta].$
We have $\mathcal{V}\left(U^{i},H\right)\left[k\delta,\left(k+1\right)\delta\right]=$
\begin{multline*}
\left(v+C_{H}\right)\bar{T}_{A}+p\mathcal{V}\left(U^{i}-\delta,N\right)\left[\left(k-1\right)\delta,k\delta\right]\\
+\left(1-p\right)\mathcal{V}\left(U^{i}-\delta,H\right)\left[\left(k-1\right)\delta,k\delta\right].
\end{multline*}
A bit of algebra gives $\mathcal{V}(U^{i},H)[k\delta,\left(k+1\right)\delta]=\delta_{1}^{D}+\left(1-p\right)\mathcal{V}\left(U^{i}-\delta,H\right)\left[\left(k-1\right)\delta,k\delta\right],$
if $U^{i}\leq\omega+\delta,$ and $\mathcal{V}(U^{i},H)[k\delta,\left(k+1\right)\delta]=\delta_{1}^{D}+\mathcal{V}\left(U^{i}-\delta,H\right)\left[\left(k-1\right)\delta,k\delta\right]-p\lambda_{N}^{D}\bar{T}_{A},$
otherwise. Solving this recursive equation for the case of $U^{i}\leq\omega+\delta$
gives $\mathcal{V}(U^{i},H)[k\delta,\left(k+1\right)\delta]=$ 
\begin{multline}
\delta_{1}^{D}+\delta_{1}^{D}\left(1-p\right)+\ldots+\delta_{1}^{D}\left(1-p\right)^{k-1}\\
+\left(1-p\right)^{k}\mathcal{V}\left(U^{i}-\delta k,H\right)\left[0,\delta\right].\label{eq:deriveUless}
\end{multline}
Using initial condition $\mathcal{V}(U,H)[0,\delta]=U\chi_{H}^{D}$
produces $f^{D}(U^{i})$ for $U^{i}\leq\omega.$ For $U^{i}>\omega+\delta,$
consider the integer $k_{1}$ such that $(k-k_{1}-1)\delta\leq\omega<(k-k_{1})\delta.$
Then 
\begin{multline*}
\mathcal{V}(U^{i},H)[k\delta,\left(k+1\right)\delta]=k_{1}\left(\delta_{1}^{D}-p\lambda_{N}^{D}\bar{T}_{A}\right)\\
+\mathcal{V}\left(U^{i}-k_{1}\delta,H\right)\left[(k-k_{1}-1)\delta,(k-k_{1})\delta\right].
\end{multline*}
But the last term is simply $f^{D}\left(U^{i}-k_{1}\delta\right),$
and $k_{1}=k\left[U^{i}-\omega\right]$ defined in Eq. (\ref{eq:defK}).
Substituting from Eq. (\ref{eq:deriveUless}) gives the entire function
$f^{D}(U^{i}),$ $U^{i}\in\mathbb{U}.$

\section{Derivation of $k[\omega]$ and $\omega$}

\label{sec:proofOmega}

We solve first for $k[\omega]$ and then for $\omega.$ Because of
the floor function in $k[\omega],$ we have that $\omega\in\left[k\left[\omega\right]\delta,\left(k\left[\omega\right]+1\right)\delta\right).$
Then for some $\epsilon\in[0,1),$ $\omega=\left(k\left[\omega\right]+\epsilon\right)\delta.$

Note that $f^{D}(\omega)=\bar{T}_{A}\lambda_{N}^{D},$ \emph{i.e.},
the expected gain of surveillance is equal to the security risk at
$U^{i}=\omega.$ Therefore, we have $\bar{T}_{A}\lambda_{N}^{D}=$
\begin{equation}
\chi_{H}^{D}\left(\omega-\delta k[\omega]\right)\left(1-p\right)^{k[\omega]}+\frac{\delta_{1}^{D}}{p}\left(1-(1-p)^{k[\omega]}\right).\label{eq:riskEqRewardDerive}
\end{equation}
Substituting for $\omega,$
\begin{multline*}
\bar{T}_{A}\lambda_{N}^{D}-\frac{\delta_{1}^{D}}{p}=\left(k\left[\omega\right]+\epsilon\right)\delta\chi_{H}^{D}\left(1-p\right)^{k[\omega]}\\
-\delta k[\omega]\chi_{H}^{D}\left(1-p\right)^{k[\omega]}-(1-p)^{k[\omega]}.
\end{multline*}
This reduces to 
\[
\bar{T}_{A}\lambda_{N}^{D}-\frac{\delta_{1}^{D}}{p}=\epsilon\delta\chi_{H}^{D}\left(1-p\right)^{k[\omega]}-(1-p)^{k[\omega]},
\]
which is uniquely solved by the $k[\omega]$ in Theorem \ref{thm:omegaClosedForm}.
Now solving Eq. (\ref{eq:riskEqRewardDerive}) for $\omega$ obtains
the result in Lemma \ref{thm:omegaClosedForm}.

\section{Derivation of $\bar{\mathcal{V}}_{\theta^{*}}(\bar{T}_{A})$}

\label{sec:proofWorstVal}

We solve the value function in two cases.

\subsection{Limit as $\bar{T}_{A}\to0$}

As $\bar{T}_{A}\to0,$ $\omega$ and $\delta$ decrease, so $U^{0}>\omega+\delta,$
and the value functions follow $f_{2}^{D}.$ Therefore, we find the
limit of $f_{2}^{D}$ as $\bar{T}_{A}\to0.$ As $\bar{T}_{A}\to0,$
$k[U^{0}]-k_{1}[U^{0}]$ remains finite, but $\delta_{1}^{D}\to0,$
and $\delta k[U^{0}]$ approaches $U^{0}.$ Therefore, the first two
terms of $f_{2}^{D}$ approach zero. The last term expands to 
\[
\bar{T}_{A}\left\lfloor \frac{U^{0}-\omega}{v\bar{T}_{A}}\right\rfloor \left(v+C_{H}+C_{N}\frac{p}{1-p}\right).
\]
As $\bar{T}_{A}\to0,$ this approaches 
\begin{equation}
U^{0}\left(1+\frac{1}{v}\left(C_{H}+C_{N}\frac{p}{1-p}\right)\right).\label{eq:proofValTaToZero}
\end{equation}
Now, manipulation of Eq. (\ref{eq:totDefUtil}) yields 
\[
\mathcal{V}_{\theta^{\dagger}}\left(\bar{T}_{A}\right)=f_{2}^{D}\left(U^{0}\right)+\bar{T}_{A}C_{N}\frac{p}{1-p}.
\]
But as $\bar{T}_{A}\to0,$ the second term approaches zero. Thus $\mathcal{V}_{\theta^{\dagger}}(\bar{T}_{A})$
approaches Eq. (\ref{eq:proofValTaToZero}). We have proved Eq. (\ref{eq:valLowComb}).

\subsection{Large $\bar{T}_{A}$ }

There are several cases. First, consider $\delta<\omega$ and $\bar{T}_{A}\geq U^{0}/v.$
The second condition implies that $D$ keeps $A$ in the first honeypot
that he enters until all residual utility is exhausted, which produces
utility $(v+C_{H})U^{0}/v$. The first condition implies that $U^{0}/v>\bar{T}_{\omega},$
so $\bar{T}_{A}>\bar{T}_{\omega},$ which means that $D$ ejects $A$
from the first normal system that he enters, which produces $0$ utility.
The weighted sum of these utilities gives Eq. (\ref{eq:valHighComb}).

Next, consider  $\delta>\omega$ and $\bar{T}_{A}\geq U^{0}/\omega.$
The first condition implies that $U^{0}/v<\bar{T}_{\omega},$ so it
not guaranteed that $\bar{T}_{A}\geq\bar{T}_{\omega}.$ But if $\bar{T}_{A}\geq\bar{T}_{\omega},$
$D$ ejects $A$ from the first normal system that he enters, and
we have Eq. (\ref{eq:valHighComb}).

\bibliographystyle{plain}
\bibliography{jpDecepBib}

\end{document}